\documentclass[12pt]{article}
\usepackage{epsfig,epsf,amssymb,latexsym}
\usepackage{color}
\usepackage{multicol}
                                                                                                 
\begin{document}
                                                                                                 
\title{\bf Toward a Quantum theory of Gravity and a Resolution of the Time paradox}
\author{Edward Tetteh-Lartey\thanks{lartey@fnal.gov} \\
Department of Physics, Texas $A\&M$ University, College Station, TX 77845,
USA
}
                                                                                                 
\date{\today}
                                                                                                 
\maketitle
                                                                                                 
\small{
One of the major issues confronting theoretical physics is finding a quantum theory of gravity and a resolution to the cosmological constant problem. It is believed that a true quantum theory of gravity will lead to a solution to this problem. Finding a quantum theory of gravity has been a difficult issue mainly because of the high energy scale required for testing quantum gravity which is far the reach of current accelerators. Also general relativity does not possess a natural time variable, thus the nature of time is not clear in quantum gravity, a problem called the time paradox.
The two main approaches to a quantum theory of gravity are string theory and loop quantum gravity. String theory unifies all interaction but provides a perturbative background dependent formulation which violates general covariance. Loop quantum gravity provides a non-perturbative approach but does not provide a unified theory of interactions, which most physicists believe should be the case at the Planck scale energies. It doesn't also seem to connect with low energy phenomena.

In this note I look at how quantum cosmology provides useful inference toward a quantum gravity theory by merging inputs from the perturbative and the non-perturbative approaches, and resolving the time paradox issue.
}
                                                                                                 
\begin{multicols}{2}
\section{Introduction}
\paragraph{}
All known fundamental theories are quantum theories except general relativity. 
It would therefore seem awkward if gravity, which couples to all other fields, should remain the only classical entity in a fundamental description. 
Also, the singularity theorems of general relativity shows the breakdown of the theory at high energies and calls for a more fundamental theory to overcome this shortcomings. Thus there is a need for a quantum theory of gravity.

One of the major current issues confronting theoretical physics is finding 
a quantum theory of gravity and a resolution to the cosmological constant problem. It is expected that a correct quantum theory of gravity should give a resolution to this problem. 

The two main approaches so far are string(M) theory and canonical quantum gravity.
String(M) theory is much more ambitious and aims at a unification of all interactions within a single quantum framework. It appears to be currently the best we have toward a quantum gravity theory. However, it has fallen short of making exact predictions of our low energy universe. In addition it is a background dependent formulation over a fixed space-time thus breaking general covariance. 

In canonical quantum gravity, an attempt is made to construct, non-perturbative quantum theory of the gravitational field on its own using standard quantization rules to the classical general relativity theory. An example of this approach is loop quantum gravity. Loop quantum gravity is a background independent approach to quantum gravity. The approach makes use of the reformulation of general relativity as a dynamical theory of connections. The main ingredient of the approach is the choice of holonomies of the connections (the loop variables) as the fundamental degrees of freedom of quantum gravity. It does not provide a unified theory of all fundamental interactions but has been extremely successful in describing Planck-scale phenomena. Its main problem is the connection with low-energy phenomena.  

The concept of general covariance or diffeomorphism invariance is at the heart of general relativity. What we have learned from general relativity is that the world is not a non-dynamic manifold with dynamic fields living over it. Rather it is a collection of dynamic fields living on top of each other, with the gravitational field being just one of them.          
General relativity is an effective theory describing classical geometry at scales below the Planck scale. 
At Planck scales the concept of classical geometry no longer makes sense and we must turn to quantum geometry.    
                                                                              
In classical general relativity, there is no background geometry. Space-time geometry is not fixed, but a dynamic variable. The same should apply in the quantum realm.
But whilst in the classical realm we can make a definite prediction of the metric, in the quantum realm the metric is a fluctuating field without any definite value and we can only give a probability interpretation. Time evolution of the metric will only give a probability amplitude and not give a specified space-time. Thus is not possible to determine whether two given nearby points on a space-time manifold are space-like separated or not. Instead, the amplitudes for predictions are sums over different metrics on manifold. Points separated by a space-like intervals in one metric could be time-like separated in another metric, that contributes just as significantly to the sum.
This idea of a dynamic metric becomes a problem in quantum field theory. For instance in quantum electrodynamics, the background Minkowski metric provides the kinematic arena on which the tensor fields $F_{\mu\nu}$ propagates. It also provides the light cones and the notion of causality. The isometrics of the Minkowski metric also allows us to construct physical quantities such as fluxes of energy, momentum, and angular momentum carried by electromagnetic waves. The geometry of the Minkowski space is fixed and insensitive of the properties of the electromagnetic field.   
Thus without a background metric, defining this physical quantities in a quantum theory becomes a problem.

Another issue in quantum gravity is the problem of time. 
In ordinary quantum theory, the presence of an external time parameter t is crucial for the interpretation of the theory: Measurement take place at a certain time, matrix elements are evaluated at fixed times, and the norm of the wave-function, is conserved in time. 
The two main methods of quantization, namely, canonical quantization method due to Dirac and Feynman's path integral method rely on a global and absolute notion of time. Thus time is part of the classical background which is needed for the interpretation of measurements. In special relativistic quantum field theories, the absolute Newtonian time can be replaced by a set of time-like parameters associated to the naturally distinguished family of relativistic inertial frames. In this way time continues to be a background parameter.
However in general relativity, time as part of space-time is a dynamic quantity and local. 

The metric itself results to be a clock, and the quantization of the metric can be understood as a quantization of time. 
Thus general relativity does not seem to possess a natural time variable, while quantum theory relies quite heavily on a preferred time. Since the nature of time in quantum gravity is not clear, the classical constraints of general relativity do not contain any time parameter, and one speaks of the time paradox$~\cite{qgrav2}$. This brings up a need to modified the concept of time at a fundamental level.

A way to go around this problem in a quantum theory is to put the system of space-time geometry and matter into two main divisions:
\begin{itemize}
\item  A system of space-time and pure geometry
\item  A system of space-time, geometry and matter\footnote{By matter I mean all fields except for the metric}
\end{itemize}

The system of pure geometry may arise in very strong gravitational regimes. In this system, matter fields barely exist since they will all convert into geometry. The wave-function of this fluctuating geometries can be estimated by a path integral over geometries and topologies as given by the Hartle-Hawking's wave-function. This wave-function should satisfy the Wheeler-DeWitt equation:
\begin{equation}
\hat{H}\Phi = 0 \label{eqnw} 
\end{equation} 

This equation is obtained from the time-dependent Schrodinger equation by putting the external time parameter t, to zero.
Thus the wave-function has no external time-dependent. The  Wheeler-DeWitt equation looks like a stationary zero-energy Schrodinger equation and needs an interpretation.
What this equation is telling us is that the concept of energy has no meaning in pure geometry. 
Thus Eq($~\ref{eqnw}$) must be the fundamental equation from which the Schrodinger equation for the quantum matter theory emerges.
This can be physically described by the quantum evolution of our universe.
We can envisage our universe emerging as a pure geometry field in space-time with no matter fields.
The wave-function obtained by a path integral over all compact complex geometries and topologies will give the probability of tunneling to the universe we observe. Matter fields only appear as a result of weaking of the gravitational coupling which may be due to symmetry breaking after the universe have tunneled into existence.  
 Matter fields emerging out of geometry bring in the concept of an external time parameter and the birth of quantum field theory.
The matter fields puts a drag on the space-time geometry reducing its quantum fluctuations. The metric becomes more stable and a perturbative expansion method over a fixed background metric gives an appropriate effective description of the system.
This situation is analogous to the D-brane system. At low energy and weak couplings D-brane fluctuations can be effectively described by the open strings living on its world volume. At strong couplings D-branes become light objects and a non-perturbative description is appropriate.

In section one I give a brief overview of string theory approach to 
quantum gravity by perturbative expansion methods over a fixed background metric. I suggest that despite criticisms of being background dependent it provides an effective description in the weak coupling regimes. 
In section 2, I look at non-perturbative formulation of quantum gravity.
I suggest that for very strong coupling or curvature, where perturbative methods are not valid and we need to use non-perturbative methods, matter fields barely exist and all that exist is geometry.  In other words geometry or gravity is the dominant force and we can neglect matter fields. In this situation the system can be described by a nonperturbative path integral over geometries and topologies.
I support this argument with the quantum evolution of our universe.             
In another paper I address the issue of the cosmological constant and suggest that the small observed value is due to 
restoration of conformal symmetry in the framework of noncritical string theory$~\cite{vsel1}$.

\section{Perturbative string theory}
\paragraph{}
In perturbative string theory one starts from the Einstein-Hilbert action,
\begin{equation}
S = \frac{1}{2G^{2}}\int d^{D}x\sqrt{g}R \label{str2}
\end{equation}
 
The metric is then perturbed over a non-dynamic fixed background metric, i.e either a flat Minkowski metric or a curved metric in the case non-linear sigma model.
\begin{equation}
g_{\mu\nu}(x) = \eta_{\mu\nu} + \sqrt{G}h_{\mu\nu}(x), \label{st1}
\end{equation}

where $\eta_{\mu\nu}$ is a background kinematic metric, often chosen to be flat, G is Newton's constant, and $h_{\mu\nu}$ is a dynamic field which measures the deviation of the physical metric from the chosen background. The field $h_{\mu\nu}$ is identified as the spin 2 graviton, the quantum that is exchanged in gravitational interactions, and it is the only field that is quantized, leaving the background metric unquantized. 

The fixed background thus provides an arena for quantum fields including the graviton to be defined and propagate. And physical notions such as causality, time, and scattering states could be defined. 
The problem with this perturbative approach is $~\cite{qgl}$:
\begin{itemize}
\item Violation of Background Independence

The split of the metric distinguishes the Minkowski metric among all others reintroducing background dependence violating a key feature in Einstein's theory.
\item Violation of Diffeomorphism Covariance

The split of the metric is not diffeomorphism covariant. The diffeomorphism group is broken down to Poincare group. Noting that diffeomorphism is a local guage symmetry equivalence for general relativity, and violating of fundamental local guage symmetries is usually considered as a bad feature in Yang-Mills theories on which all other interactions are based. Background independence and violation of general covariance are thus synonymous.

\item Gravitons and Geometry

The idea of the gravitational interaction as a result of graviton exchange on a background metric contradicts Einstein's original and fundamental idea that gravity is geometry and not a force in the usual sense. This makes the perturbative description of the theory an unnatural setting and can have at most a semi-classical meaning when the metric fluctuations are very tiny.

\item Gravitons and Dynamics

In classical general relativity the metric evolves in time in an interplay with the matter present. An initially Minkowski metric can evolve to something that is far from Minkowskian at other times. In such situations the assumptions being made that h is small as compared to $\eta$ is just not dynamically stable.
\end{itemize}

In addition to this problems perturbative string theory is not Borel-summable. It is expected that a truly fundamental theory should be finite.

For the above reasons perturbative string theory has been criticized for not giving a truly fundamental theory of gravity and there is a call for a non-perturbative background independent formulation of quantum gravity. 
But here I will like to point out that in the limit of weak gravitational coupling, 
there is a system of matter and geometry. The matter fields put a drag on the space-time geometry reducing its fluctuations and thus nearly stabilizing it. The system can thus be effectively described by a perturbative expansion over a fixed background space-time.  

\section{Non-perturbative quantum gravity}
\paragraph{}
In a truly background independent formulation, no reference to any classical metric should enter neither the definition of the state space nor the dynamical variables of the theory. Rather the metric should appear as an operator allowing for quantum states which may themselves be superpositions of different background.
In a quantum theory the time evolution of a state $\Phi(x,t)$ is given by the Schroedinger equation
\begin{equation}
i\hbar\frac{d}{dt}\Phi(t)=\hat{H}\Phi(t) \label{eq1}
\end{equation}                 
where the Hamiltonian refers to evolution, as would be measured by an external observer.                                                                                                
In a closed universe where there is no external observer and thus no external time t with which leads to the quantum constraint equation, Eq($~\ref{eqnw}$).
The wave-function is given by a gravitational propagator from an initial state geometry $g^{i}$ to a final state $g_{f}$
\begin{equation}
Z(g_{i},g_{f})=\sum_{topologies}\int_{Geom(M)}dg_{\mu\nu}e^{iS[g_{\mu\nu}]} \label{eqn3}
\end{equation}

This full quantum-gravitational path integral cannot be evaluated or probably even rigorously defined, and one must resort to approximations, either semi-classical or minisuperspace approximations or a combination of both. In the minisuperspace approximation integration has to be performed over complex metrics to guarantee convergence.                                                                                                 
Evaluating this integral is extremely difficult and a major issue in quantum gravity, since 
perturbative evaluation is non-renormalizable and need to resort to non-perturbative methods.
In the Hartle-Hawking no-boundary proposal, the wave function is expressed as a path integral over compact Euclidean geometrics bounded by a given 3-geometry g.
\begin{equation}
\phi(g) =\int^{g}e^{-S_{E}} \label{eq4}
\end{equation}

But the integral above is badly divergent. This problem is rectified by additional contour rotations, extending the path integral to complex metrics. However the space of complex metrics is very large with no obvious choice of integration contour as the preferred one. In practice one assumes that the dominant contribution to the path integral is given by the stationary points of the action and evaluates $\phi_{HH}$ simply as $\phi_{HH}\sim e^{-S_{E}}$.
In a simple model, $S_{E} = -3/8\rho_{v}$ and the nucleation probability is given by:
\begin{equation}
{\it P} \sim e^{3/8\rho_{v}} \label{eqn5} 
\end{equation}

The wave-function as evaluated above gives a quantum evolution of a classical space-time and should only be seen as a close approximation to correct quantum gravitational wave-function which should be evaluated over deformed space-time metrics, i.e with non-commutative geometry.  
 One could see from $Eq(~\ref{eqn5})$ that a flat 10D compact universe with zero vacuum energy has the highest tunneling probability. The size of this space-time is small but non-zero since a zero size will have zero probability. 

\section{Discussion}
\paragraph{}
{\bf Effects of matter on geometry and vice-versa:}

General relativity describes an interaction of matter with geometry on large scales. Matter curves geometry, and gravitation is the effects of this curvature.
An initial almost Minkowski metric on space-time can evolve in time to something far from Minkowskian. The evidence of this is shown by the big bang in cosmology. Thus the metric of space-time is not fixed but a dynamic field. Could this dynamism be the response of the metric to the Liouville field or matter fields 
fluctuating in and out of the metric?.
In any case these description of matter interactions with the metric is a classical description. In the quantum realm, things may not be exactly the same. It is possible that the matter fields also cause a drag on the space-time metric reducing its quantum fluctuations. Note that in the quantum realm the fluctuations are the main dynamics of the metric field. 
 
The main question I have is does matter only curve geometry?. What other effects does it have on geometry in the classical or quantum realm?. Also are there any effects of geometry on matter?.
The above issues calls for a full investigation of matter interactions with geometry                                                                      
\section{Conclusion}
 \paragraph{}                                   
Due to lack of experimental input to test quantum gravity inference should come from quantum cosmology. Using input from quantum cosmology, I have looked at perturbative and nonperturbative approaches to a theory of quantum gravity. I suggest that both approaches are valid if the coupling is specified appropriately. 
Geometry and matter interactions can be viewed as a system of interacting fields. At strong gravitational couplings matter fields barely exist, all that exist is pure geometry. A non-perturbative approach by integrating over all metrics and topologies is valid in this regimes. 
At weak gravitational coupling, matter fields emerge out of geometry and put a drag on space-time geometry, reducing its fluctuations and hence stabilizing it. A perturbative string theory approach in which the metric is expanded over a fixed background metric is appropriate and provides an effective description of the system.
This conjecture is supported with the quantum evolution of our universe. We can envisage our universe emerging as a stringy geometrical field of deformed space-time. The tunneling probability can be calculated as a path integral over all complex compact 10D deformed geometries. The Hartle-Hawking wave-function evaluated over complex compact classical geometries only gives an approximation.
 The tunneling probability predicts a flat compact universe as the most likely to tunnel. 

I postulate that a flat 10D compact universe with all symmetries fully intact tunneled into existence at the beginning of our universe. The space-time might have been a deformed quantum space-time which gradually smoothened out and became a classical space-time as the quantum fluctuations reduced. The subsequent evolution of this classical space-time is given by classical general relativity. The matter fields we observe emerged out of geometry on this space-time when the gravitational coupling became weak. 
Weaking of the gravitational coupling could be a result of 
spontaneous breaking of a symmetry which we have not yet discovered. Further work needs to be done in this area.

I conclude that the fundamental quantum theory must be a gravitational/geometric theory with a wave-function satisfying the DeWitt-Wheeler equation. Matter fields emerge from geometry when the gravitational coupling becomes weak. This matter fields bring in the concept of an external time parameter and the birth of quantum field theory satisfying the Schroedinger's equation in the non-relativistic limit. The external time parameter t, may be the Liouville field, a field having the same signature as the intrinsic time of the space-time metric. 

Another important issue which was not discussed here but worth mentioning is the cosmological constant problem. I suggest that it could be resolve using conformal symmetry in the framework of non-critical string theory. This is discussed in detail in$~\cite{vsel1}$.

\end{multicols}
                                                                                                 

\begin{thebibliography}{99}

\bibitem{qgrav2}B. Fauser, J. Tolksdorf, E. Zeidler, ``Quantum Gravity, Mathematical Models and Experimental Bounds''.

\bibitem{qgl}D. Giulini, C. Kiefer, C. lammerzahl, ``Quantum Gravity from Theory to Experimental search, Lecture Notes in Physics''.

\bibitem{vsel2}E. Tetteh-Lartey, ``A New Perspective to Cosmic Evolution and Vacuum Selection on a Superspace'', arXiv:0705.3489. 
                            
\bibitem{vsel1}E. Tetteh-Lartey, ``Vacuum Selection on the String Landscape'', Phy. Rev D${\bf 75}$.                              

\end{thebibliography}
\end{document}